\def \SAIT #1 #2 {{\em Mem.\ Soc.\ Astron.\ It.\/} {\bf #1}, #2} 
\def \MESS #1 #2 {{\em The Messenger\/} {\bf #1}, #2} 
\def \ASTRNACH #1 #2 {{\em Astron. Nach.\/} {\bf #1}, #2} 
\def \AAP #1 #2 {{\em Astron. Astrophys.\/} {\bf #1}, #2} 
\def \AAL #1 #2 {{\em Astron. Astrophys. Lett.\/} {\bf #1}, L#2} 
\def \AAR #1 #2 {{\em Astron. Astrophys. Rev.\/} {\bf #1}, #2} 
\def \AAS #1 #2 {{\em Astron. Astrophys. Suppl. Ser.\/} {\bf #1}, #2} 
\def \AJ #1 #2 {{\em Astron. J.\/} {\bf #1}, #2} 
\def \ANNREV #1 #2 {{\em Ann. Rev. Astron. Astrophys.\/} {\bf #1}, #2} 
\def \APJ #1 #2 {{\em Astrophys. J.\/} {\bf #1}, #2} 
\def \APJL #1 #2 {{\em Astrophys. J. Lett.\/} {\bf #1}, L#2} 
\def \APJS #1 #2 {{\em Astrophys. J. Suppl.\/} {\bf #1}, #2} 
\def \APSS #1 #2 {{\em Astrophys. Space Sci.\/} {\bf #1}, #2} 
\def \ASR #1 #2 {{\em Adv. Space Res.\/} {\bf #1}, #2} 
\def \BAIC #1 #2 {{\em Bull. Astron. Inst. Czechosl.\/} {\bf #1}, #2} 
\def \JSQRT #1 #2 {{\em J. Quant. Spectrosc. Radiat. Transfer\/} {\bf #1}, #2} 
\def \MN #1 #2 {{\em Mon. Not. R. Astr. Soc.\/} {\bf #1}, #2} 
\def \MEM #1 #2 {{\em Mem. R. Astr. Soc.\/} {\bf #1}, #2} 
\def \PLR #1 #2 {{\em Phys. Lett. Rev.\/} {\bf #1}, #2} 
\def \PASJ #1 #2 {{\em Publ. Astron. Soc. Japan\/} {\bf #1}, #2} 
\def \PASP #1 #2 {{\em Publ. Astr. Soc. Pacific\/} {\bf #1}, #2} 
\def \NAT #1 #2 {{\em Nature\/} {\bf #1}, #2} 

\documentstyle{memsait} 
\input epsf.sty 
\begin{opening} 
\title{The Multiplicity Function of groups of galaxies from CRoNaRio catalogues} 
\author{E. De Filippis$^1$, G. Longo$^2$, S. Andreon$^2$, R. Scaramella$^3$, Testa V.$^3$, R.de Carvalho$^4$, G.Djorgovski$^5$} 
\institute{$^1$Universit\`a Federico II, Naples, Italy\\ 
$^2$Osservatorio Astronomico di Capodimonte, Napoli, Italy\\ 
$^3$Osservatorio Astronomico di Monte Porzio, Roma, Italy\\ 
$^4$Observatorio Nacional, Rio de Janeiro, Brazil\\
$^5$Department of Astronomy, Caltech, USA}
\date{} 
\end{opening}

\begin{document} 
 
\oddpagefooter{}{}{} 
\evenpagefooter{}{}{} 
\  
 
\begin{abstract} 
The projected multiplicity function of galaxies gives (per square degree) the density of galaxy aggregates formed by N members. We use the CRoNaRio matched catalogues to derive the low N ($N<15$) multiplicity function from D-POSS data. The van Albada (1982) algorithm was implemented and used to identify candidate groups in the CRoNaRio catalogues. In absence of redshift surveys complete down to the magnitude limit of the DPOSS material and covering a wide enough area, the performances of the algorithm were tested on realistically simulated catalogues.\\
The application to a set of 13 CRoNaRio catalogues allowed us to derive a multiplicity function as accurate as those available in literature obtained from redshift surveys.  
\end{abstract} 
 
\section{Introduction} 
The multiplicity function (hereafter MF) is a powerful tool to test the various cosmological scenarios (Combes \& Boiss\'e 1991) and provides an independent way to measure the index $n$ of the initial perturbation spectrum $n$ at the era of baryonic recombination (Turner, Gott III 1976).\\ 
 The CRoNaRio project is a joint enterprise among Caltech and the astronomical observatories of Napoli, Roma and Rio de Janeiro, aimed to produce the first general catalogue of all objects visible on the DPOSS (Digitised Palomar Sky Survey). The final Palomar-Norris catalogue will include astrometric, photometric (in the three Gunn-Thuan bands g, r and i) and rough morphological information for an estimated $2 \times 10^9$ stars and $5 \times 10^7$ galaxies. More than $60 \%$ of the catalogues are already available and are currently being used for many scientific applications.\\ 
 
\section{The van Albada algorithm} 
\begin{figure} 
\hspace{3.0cm} 
\epsfxsize=9.0cm 
\epsfbox{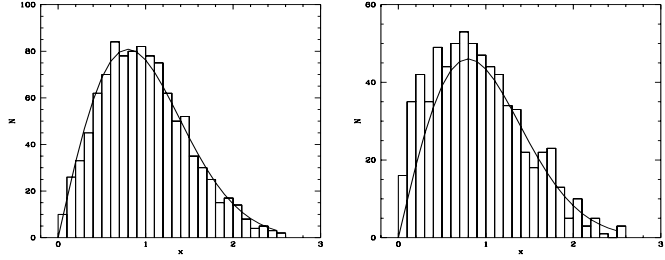} 
\caption[h] {Expected (left) and observed (right) distribution of $x$.} 
\label{fig:poisson} 
\end{figure}

In order to compile a catalog of candidate groups of galaxies in absence of redshift information, we implemented a slightly modified version of the van Albada algorithm (Soares 1989).\\ 
This algorithm makes use of apparent magnitude and projected position in the sky only, and gives, for each pair of adjacent galaxies, their probability of being physically related.\\ 
Assuming a Poisson statistic, the probability that the angular distance of a fixed galaxy to its nearest non-physical companion lies between $\theta$ and $\theta+d\theta$ is given by: 
$$P_1(\theta)d\theta=exp\left[ -\pi \theta ^2 n \right] \cdot 2 \pi \theta n d\theta.$$ 
The introduction of an adimensional distance $x$ (defined as the ratio of the observed to the expected mean distance to the nearest neighbour) allows to combine the angular separation of different pairs into a single distribution (Fig. \ref{fig:poisson}) removing the effects of clustering in the background.\\ 
 
\section{The accuracy of the algorithm} 
In order to test the accuracy (lost groups) and the reliability (spurious groups) of the algorithm, we tested it on $70$ realistically simulated sky fields.\\ 
We first produced the galaxy background by assuming uniform distribution and the field luminosity function given by Metcalfe et al. 1994; then we added simulated groups of galaxies according to the multiplicity function by Turner \& Gott III 1976, with redshift computed according to: 
$$N(z)=4 \pi \frac{\rho_0}3 \left[ \frac 8 {H_0^3} \left( z+1- \sqrt{1+z} \right)^3 \right]$$ 
and absolute magnitude of the brightest galaxy in the group taken from the cumulative luminosity function for groups of galaxies: 
$$\Phi(M)dM=\Phi^* \left[ 10^{0.4} \left( M^*-M \right) \right]^{\alpha+1} \cdot exp \left[ -10^{\left( 0.4 \left( M^*-M \right) \right) } \right] dM$$ 
where $M^*=-20.85$ and $\alpha= -0.83\pm 0.17$. 
Other parameters which were varied in the course of the simulations were: 
\begin{itemize} 
\item the diameter of the group inside a Gaussian distribution centered at: $D_0=0.26Mpc$; 
\item the maximum possible redshift for a group, which were assumed to fall in the range: $0.2 \leq z \leq 0.4$. 
\end{itemize} 
 
\begin{figure} 
\epsfxsize=12.0cm 
\hspace{3.0cm} 
\epsfbox{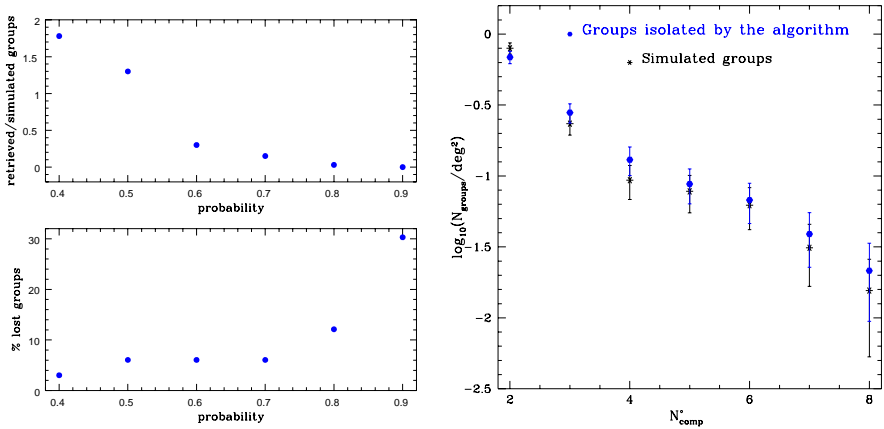} 
\caption[h] {Number of spourious and lost groups vs the lower limit of the probability (left) and the multiplicity function of simulated groups (right).} 
\label{fig:prob} 
\end{figure}

Different simulations were performed in order to optimize the lower limit of the probability for which two galaxies were considered as physical companions by the algorithm. The optimal value, id est the value ensuring the best compromise between accuracy and reliability (Fig. \ref{fig:prob}), 
turned out to be $p \geq 0.6$.\\ 
With this choice, the modified Van Albada algorithm succeds in retrieving, from the simulated catalogues, more than $90 \%$ of the groups having more than three components.\\ 
\vspace{0.0cm} 
 
\section{A preliminary Multiplicity Function from CRoNaRio data} 
\begin{figure} 
\epsfxsize=6.5cm 
\hspace{3.5cm} 
\epsfbox{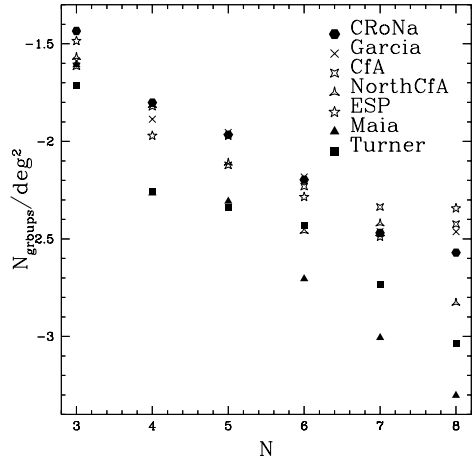} 
\caption[h]{MF of CroNaRio data compared to MFs in literature; different surveys have been normalized by their galaxy density}
\label{fig:disc} 
\end{figure}

The figure (Fig. \ref{fig:disc}) gives the MF obtained from $13$ CRoNaRio plates (subtending a total solid angle of $483 \: deg^2$). In the same figure we also compare it with six other MFs taken from literature: 
\begin{enumerate} 
\item {\bf MF derived from magnitude limited surveys making use of redshift information:} {\bf Garcia} (Garcia et al. 1993), {\bf CfA} (Geller \& Huchra 1983), {\bf NorthCfA} (Ramella, Pisani 1997), {\bf ESP} (Ramella et al. 1998)
\item {\bf MF derived from diameter limited survey: Maia} (Maia \& Da Costa 1989)
\item {\bf MF derived from limited magnitude survey without redshift information: Turner} (Turner \& Gott 1976)
\end{enumerate} 
Agreement is found between our FM, obtained with our algorithm without any tridimensional data, and the ones in literature using redshift information. Therefore, with our algorithm, even in absence of redshift information, it is possible to explore wide areas of the sky and to retrieve groups ($N>2$) with an efficiency comparable to that of 3-D surveys.\\

\section{An estimate of $n$}
For the groups with at least one redshift available, MF (as a function of luminosity) was calculated. With a fit of the observed data to the theoretical function
\begin{equation}
f(L)=1-\pi^{-1/2} \int_{\left( \frac L {L_c} \right) ^{\left( 1+n_{oss}/3 \right )}} ^{\infty} e^{-t} t^{\frac 1 2 -1} dt.
\end{equation}
we obtained: $\: \: \: n_{oss}=-1.0\; (-1.2,\; -0.8); \: \: \: c=L_c/L^*=3.5\; (3.2,\; 3.8).$\\
This allows an estimate of the spectral index $n$: 
$$n=1.5+1.5n_{oss}=0.0\; (-0.3,\; 0.3).$$

\end{document}